\documentstyle[12pt]{article}
\newcommand{\ie} {{\it i.e.}}

\def\be{\begin{eqnarray}}
\def\ee{\end{eqnarray}}

\def\half{{\textstyle{1 \over 2}}}

\begin{document}
\begin{flushright} USITP-97-02\\ February 1997
\end{flushright}

\bigskip
\Large
\begin{center}
\bf{Tensionless p-branes with manifest conformal
 invariance}
\bigskip

\normalsize 
\bigskip

P. Saltsidis\footnote{e-mail 
address:panos@vanosf.physto.se}\\  {\it
ITP\\ University of Stockholm\\
 Box 6730, Vanadisv\"agen 9\\ S-113 85
Stockholm\\  SWEDEN}\\
\end{center}
\vspace{2.0cm}
\normalsize
\bigskip
\bigskip

{\bf Abstract:}
The quantization of the tensionless p-brane is
discussed. Inspection of the constraint algebra
reveals that the central extensions for the p-branes
have a  simple form.  Using a Hamiltonian BRST scheme we
find that the quantization is consistent in any
space-time dimension while the quantization of the
conformal tensionless p-brane gives a critical dimension
$d=2$.

\eject
\section{Introduction}
The quantum behavior of the
tensionless string is found to be very different  from
that of the usual string, since quantization does not
give rise to a critical dimension \cite{liraspsr}.
However, the tensionless string is space-time
conformally invariant \cite{jiulbs,BigT}. It is this
symmetry that replaces the Weyl invariance in the $T\to
0$ limit. Requiring this symmetry to be a fundamental
symmetry of the theory led  to the construction of the
conformal string, a string with manifest space-time
conformal symmetry
\cite{ISBE,us}. Using a Hamiltonian BRST scheme this
model was quantized and a critical dimension $d=2$ was
found. The mass spectrum of this anomaly free theory was
investigated in 
\cite{spectrum}. There a
BRST treatment of the physical states revealed that the
string collapses to a massless particle, a result which
agrees with the classical treatment.

In this paper we investigate the quantum properties of 
the generalization of the conformal strings to higher
dimensional objects, the tensionless p-branes with
manifest conformal space-time symmetry. The
quantization of the usual p-branes is a very complicated
problem because the theory is highly nonlinear. The
constraint algebra is of rank greater than one, which
means that the structure coefficients are field
dependent. Therefore  a quantum mechanical discussion of
this algebra is difficult. However consistent
quantization may be carried out for the tensionless
p-brane.  As a result the corresponding theory is found
to be anomaly free for  arbitrary space-time
dimension $d$. However when the space-time conformal
symmetry is investigated a critical dimension
$d=2$ is again found.

The content of the paper is as follows: In Section
\ref{pbrane}  we discuss the tensionless p-brane. An
 analysis of the constraint algebra reveals a simple
form for the central extensions. Investigating the
quantum behavior  we find the theory to be anomaly
free. Using normal ordering we reproduce the result of
\cite{umms} which  puts restrictions on the space time
dimensions for the tensile p-brane. We also make  some
remarks on the generalization of this result to the
case of the spinning p-brane. In Section \ref{conbrane}
we proceed to the discussion of the tensionless
conformal p-brane. Using the techniques of \cite{us} we
recover a consistent theory in the extended phase space
of ghost and matter fields, in two space-time dimensions.

\section {The tensionless p-brane} \label{pbrane}

The action for the tensionless p-brane is given by
\cite{HLU}
\be   
S=\int d^{p+1}\sigma
V^{a}V^{b}\gamma_{ab}\label{action},
\ee   
where $\gamma_{ab}=\partial_{a}
X^{\mu}\partial_{b}X_{\mu}$ is the induced metric
$\mu =0,1$ is a space-time index,  $a=0,\ldots, p$
 is a world $(p+1)$-volume index and 
$V^{a}$ is a weight
$w=-\half$ contravariant $p$-dimensional vector density.
It is a generalization of the action of the tensionless
string first given in \cite{ulbsgt1}. The action is
invariant under
 world-sheet
 diffeomorphisms
\be
\delta_{\epsilon}X^{\mu}&= &\epsilon\cdot\partial 
X^{\mu},\nonumber\\
\delta_{\epsilon}V^{a}&=&
-V\cdot\partial\epsilon^{a}+\epsilon\cdot\partial 
V^{a}+\half (\partial\cdot\epsilon )
V^{a}.\nonumber
\ee  

Passing to the Hamiltonian formulation we  find, 
\cite{us}, that the
Hamiltonian is a linear combination of the $p+1$ 
constraints
\be
\phi^{-1}(\sigma_1,\ldots,\sigma_p)
&=&P^{\mu}P_{\mu}(\sigma_1,\ldots,\sigma_p)=0\nonumber\\
L^{\alpha}(\sigma_1,\ldots,\sigma_p)&=&P^{\mu}\partial
_{\alpha}X_{\mu} (\sigma_1,\ldots,\sigma_p)=0,\nonumber
\ee
 which is expected since the theory is reparametrization
 invariant.  Note that  the greek index $\alpha$ runs 
from 1 to $p$. In Fourier modes the 
constraints read\footnote{For simplicity we are 
considering 
closed p-branes.}
\be
\phi^{-1}_{m_1,\ldots ,m_p}&=&\half
\sum_{k_1,\ldots, k_p =-\infty}^{+\infty}p_{m_1 -
k_1,\ldots, m_p - k_p}
\cdot p_{k_1,\ldots ,k_p}=
0,\label{ccon1}\\
L^{\alpha}_{m_1,\ldots
,m_p}&=&-i\sum_{k_1,\ldots, k_p
=-\infty}^{+\infty} 
 k_{\alpha}p_{m_1 -
k_1,\ldots, m_p - k_p}\cdot x_{k_1,\ldots ,k_p}=
0\label{ccon2}
\ee  
and they satisfy the following algebra
\be
\left[
{\phi}^{-1}_{m_1,\ldots ,m_p},{L}^{\alpha}_{n_1,\ldots
,n_p}\right]   &=&
(m_{\alpha}-n_{\alpha}){\phi}^{-1}_{m_1
+n_1,\ldots,m_p+n_p},\label{excont}\\
  \left[ {L}^{\alpha}_{m_1,\ldots
,m_p},{L}^{\beta}_{n_1,\ldots ,n_p}\right]
&=&m_{\beta}L^{\alpha}_{m_1
+n_1,\ldots,m_p+n_p}- n_{\alpha}L^{\beta}_{m_1
+n_1,\ldots,m_p+n_p}\nonumber\\
&&+A^{\alpha\beta}(m_1,\ldots
,m_p)\delta_{m_1+n_1}\ldots\delta_{m_p+n_p},
   \label{d12} 
\ee  
where the basic non zero commutators are
\be
\left [x^{\mu}_{m_1,\ldots ,m_p},p^{\nu}_{n_1,\ldots
,n_p}\right ]
=i\delta_{m_1+n_1}\ldots\delta_{m_p+n_p}
\eta^{\mu\nu}.\nonumber
\ee  

The right hand side of equation (\ref{d12}), when $m_1
+n_1 = \ldots = m_p + n_p =0$, is expressed in terms of
$L^{\alpha}_{0,\ldots,0}$. But this operator is not well
defined since it depends on the different orderings of
$x^{\mu}_{m_1,\ldots ,m_p}$ and $p^{\mu}_{m_1,\ldots ,m_p}$.
Taking into account this ambiguity we include possible
central extensions in the right hand side of the
commutators (\ref{d12}). The values of these central
extensions are constrained by the Jacobi identities 
\be
&&\left [ \left[ {L}^{\alpha}_{m_1,\ldots
,m_p},{L}^{\beta}_{n_1,\ldots ,n_p}\right],
{L}^{\gamma}_{k_1,\ldots ,k_p}\right]+
\left [ \left[ {L}^{\gamma}_{k_1,\ldots
,k_p}, {L}^{\alpha}_{m_1,\ldots
,m_p}\right],
{L}^{\beta}_{n_1,\ldots ,n_p}\right]+\nonumber\\
&&\left [
\left[ {L}^{\beta}_{n_1,\ldots
,n_p},{L}^{\gamma}_{k_1,\ldots ,k_p}\right],
{L}^{\alpha}_{m_1,\ldots ,m_p}\right] =0\nonumber
\ee
and the commutator relation
\be
\left[ {L}^{\alpha}_{m_1,\ldots
,m_p},{L}^{\beta}_{n_1,\ldots ,n_p}\right]=-
\left[{L}^{\beta}_{n_1,\ldots ,n_p},
{L}^{\alpha}_{m_1,\ldots ,m_p}\right]\nonumber.
\ee
We find that for $p>1$
\be
A^{\alpha\beta}(m_1,\ldots
,m_p)=A^{\alpha\beta}(m_{\alpha},m_{\beta})=\half\left
(m_{\beta}d^{\alpha}+m_{\alpha}d^{\beta}\right
),\nonumber
\ee
where $d^{\alpha}$ are constants. 

In order to clarify
the implications of the last relation  we take
$\alpha=\beta$ in (\ref{d12}). Then
\be
\left[ {L}^{\alpha}_{m_1,\ldots
,m_p},{L}^{\alpha}_{n_1,\ldots ,n_p}\right]
&=&(m_{\alpha}-n_{\alpha})L^{\alpha}_{m_1
+n_1,\ldots,m_p+n_p}+m_{\alpha}d^{\alpha}
\delta_{m_1+n_1}\ldots\delta_{m_p+n_p}.\nonumber
\ee
The corresponding relation for the string ($p=1$) is
\be
\left[ {L}_{m},{L}_{n}\right] &=&(m-n)
    {\phi}^{L}_{m+n}+({d}_{3}m^{3}+{d}_{1}m)\delta_{m+n}.
   \nonumber
\ee
Thus we find that in the case of the tensionless p-branes
with
$p>1$, the central extensions in the algebra of the
constraints become "smoother" since their cubic terms
have to vanish due to the Jacobi identities. It is
interesting to note that the same cancellation will also
occur in the case of the usual p-branes. The
constraints $L^{\alpha}$ are not modified by the
nonzero tension and so the results obtained here for the
subalgebra (\ref{d12}) are also valid for the tensile
p-brane.

We are going to  investigate
the quantum theory of this model within the
framework of a BRST quantization which is
considered to  be the fundamental way to quantize
general gauge theories. The BRST quantization requires
the introduction   of new operators, the  Faddeev-Popov
ghosts. To every constraint,
 one introduces a ghost pair
$c^{A}_{m_1,\ldots,m_p}$, $b^{A}_{m_1,\ldots,m_p}$,
$A\in\{-1,L_{\alpha}\}$, that is fermionic. These ghosts
satisfy the fundamental anticommutation relations
\be
\left
\{c^{A}_{m_1,\ldots,m_p},b^{B}_{n_1,\ldots,n_p}\right \}
=\delta_{m_1+n_1}\ldots\delta_{m_p+n_p}\delta^{AB}.\nonumber
\ee  
The generator of BRST transformations, 
the BRST charge \cite{FRAD} is found to be
\be
Q & = &\sum_{k_1,\ldots,k_p}
\phi_{-k_1,\ldots,-k_p}^{-1}c_{k_1,\ldots,k_p}^{-1}+
\sum_{\alpha =1}^{p}
\sum_{k_1,\ldots,k_p}{L^{\alpha}}_{-k_1,\ldots,-k_p}
c_{k_1,\ldots,k_p}^{L_{\alpha}}\nonumber\\
&&-
\sum_{\alpha
=1}^{p}\sum_{k_1,\ldots,k_p}\sum_{l_1,\ldots,l_p}
(k_{\alpha}-l_{\alpha})c_{-k_1,\ldots,-k_p}^{-1}
c_{-l_1,\ldots,-l_p}^{L_{\alpha}}
b_{k_1+ l_1,\ldots,k_p + l_p}^{-1}\nonumber\\
&&-\half
\sum_{\alpha,\beta
=1}^{p}\sum_{k_1,\ldots,k_p}\sum_{l_1,\ldots,l_p}
k_{\beta}c_{-k_1,\ldots,-k_p}^{L_{\alpha}}
c_{-l_1,\ldots,-l_p}^{L_{\beta}}
b^{L_{\alpha}}_{k_1+ l_1,\ldots,k_p +
l_p}\nonumber\\
&&+\half
\sum_{\alpha,\beta
=1}^{p}\sum_{k_1,\ldots,k_p}\sum_{l_1,\ldots,l_p}
l_{\alpha}c^{L_{\alpha}}_{-k_1,\ldots,-k_p}
c_{-l_1,\ldots,-l_p}^{L_{\beta}}
b_{k_1+ l_1,\ldots,k_p +
l_p}^{L_{\beta}}
.\label{Q}
\ee  
The classical nilpotency, ${Q}^2=0$, is guaranteed 
by construction. To check the nilpotency of the quantum
${\cal Q}=\half(Q + Q^{\dagger})$, which is constructed
to be hermitian,  we use   the
extended constraints
$\tilde{\phi}^{I}_{m_1,\ldots,m_p}$. These are  BRST
invariant extensions of  the original constraints and
they satisfy the same algebra as the original ones for
first rank systems. They are defined  by  the equation 
\be
\tilde{\phi}^{I}_{m_1,\ldots,m_p}\equiv \{
b^{I}_{m_1,\ldots,m_p},{\cal Q}\},\nonumber
\ee 
which means that
\be
\lefteqn{\tilde{\phi}^{-1}_{m_1,\ldots,m_p}=
\phi^{-1}_{m_1,\ldots,m_p}
-\sum_{\alpha
=1}^{p}\sum_{k_1,\ldots,k_p}
(m_{\alpha}-k_{\alpha})c_{-k_1,\ldots,-k_p}^{L_{\alpha}}
b_{k_1+ m_1,\ldots,k_p + m_p}^{-1},}\\
\lefteqn{\tilde{L}^{\alpha}_{m_1,\ldots,m_p}=
{L}^{\alpha}_{m_1,\ldots,m_p}+\sum_{k_1,\ldots,k_p}\left
[ (k_{\alpha}-m_{\alpha})c_{-k_1,\ldots,-k_p}^{-1}
b_{k_1+ m_1,\ldots,k_p + m_p}^{-1}\right .}\label{L}\\
&&+\sum_{\beta
=1}^{p}\left .(
k_{\alpha}c_{-k_1,\ldots,-k_p}^{L_{\beta}}
b_{k_1+ m_1,\ldots,k_p + m_p}^{L_{\beta}}-
m_{\beta}c_{-k_1,\ldots,-k_p}^{L_{\beta}}
b_{k_1+ m_1,\ldots,k_p +
m_p}^{L_{\alpha}})\right ]\nonumber.
\ee
We can now calculate the BRST anomaly using a method
described in
\cite{MarnABRST,ISBE}. There it is shown that
\be 
{\cal Q}^{2}=\half \sum_{I,J}\sum_{m_1,\ldots,m_p}{\tilde
{d}}^{IJ}_{m_1,\ldots,m_p}c^{I}_{m_1,\ldots,m_p}
c^{J}_{-m_1,\ldots,-m_p},\nonumber
\ee 
where ${\tilde {d}}^{IJ}$ are the central extensions
of the extended
 constraints  algebra.
This means that 
 \be {\cal Q}^{2} =
\half\sum_{\alpha}\tilde{d}^{\alpha}m_{\beta}
c^{L_{\alpha}}_{m_1,\ldots,m_p}
c^{L_{\beta}}_{-m_1,\ldots,-m_p}.\label{as}
\ee 
The exact values of $\tilde{d}^{\alpha}$ depend
on the  vacuum and ordering we use. The simplest
and safest method to determine
 these constants is to calculate the vacuum expectation
value of the commutators (\ref{d12})
for the extended constraints.

According to arguments presented in \cite{BigT}  the  
vacuum suitable
for  tensionless strings is not   one annihilated 
by the positive modes
of the operators but  one annihilated  by the 
momenta\footnote{cf. the
vacuum for a particle.}
\be
  p^{\mu}_{m}|0\rangle_{p} =0 \quad \forall m.\label{vac}
\ee 
In the case of the tensionless p-brane we take the
vacuum to be defined also by (\ref{vac}). The operators
$p^{\dagger}_{m_1,\ldots,m_p}$,
$x^{\dagger}_{m_1,\ldots,m_p}$ and
$p_{m_1,\ldots,m_p}$,
$x_{m_1,\ldots,m_p}$, as a consequence of the
particle-like nature of the equations of motion
\cite{BAZH}, are not connected with positive and
negative frequency parts of the field operators \ie, to
treat them as creation and annihilation operators would
be most artificial.

Following the
prescription of
\cite{MarnBRST}, we will take  the {\em ket } states to
be built from our vacuum of choice,
$|0\rangle_{p}$, and the {\em bra } states to be built 
from
$\mbox{}_{x}\langle 0|$ satisfying
$\mbox{}_{x}\langle 0|0\rangle_{p}=1$. 

For the vacuum (\ref{vac}) and from the requirement 
that the BRST
charge (\ref{Q}) should annihilate the vacuum,
 we obtain further
requirements on the ghost part of the vacuum. 
Doing
this we find that the vacuum has to
 satisfy the following conditions 
\be
 p^{\mu}_{m_1,\ldots,m_p}|0\rangle 
=b^{-1}_{m_1,\ldots,m_p}|0\rangle = 0,\nonumber\\ 
  \langle 0|x^{\mu}_{m_1,\ldots,m_p}  =\langle
0|c^{-1}_{m_1,\ldots,m_p}=0.\nonumber
\ee
The expectation value of the commutator (\ref{d12}) is
\be\label{fcon1}
\left\langle
0\left |\left[\tilde{L}^{\alpha}_{m_1,\ldots
,m_p},\tilde{L}^{\alpha}_{-m_1,\ldots
,-m_p}\right]\right |0\right\rangle
=2m_{\alpha}\left\langle
0\left
|\tilde{L}^{\alpha}_{0,\ldots,0}\right|0\right\rangle
+m_{\alpha}\tilde{d}^{\alpha}\nonumber\\
\Rightarrow
0=2m_{\alpha}a^{\alpha}_L+m_{\alpha}\tilde{d}^{\alpha}
\Rightarrow
\tilde{d}^{\alpha}=-2a^{\alpha}_L\label{alpha}
\ee
where $a^{\alpha}_L\equiv \left\langle
0\left
|\tilde{L}^{\alpha}_{0,\ldots,0}\right|0\right\rangle$.
But from (\ref{L}) for a hermitian BRST charge
${\cal{Q}}$ we will have \cite{us}
\be
a^{\alpha}_L=
-\half(d+1+p)\sum_{k_{\alpha}=-\infty}^{+\infty}
k_{\alpha}\sum_{k_1,
\ldots,k_{\alpha -1},k_{\alpha
+1},\ldots,k_p}1=0.\nonumber
\ee
Thus from the relations (\ref{as}) and (\ref{alpha}) we
deduce that the BRST charge is nilpotent for any
space-time dimension $d$, as was also observed in
\cite{BAZH} using other methods. So in the theory of
tensionless p-branes, just as in the theory of
tensionless strings the critical dimension is absent and
the theory is quantum mechanically consistent for any
dimension $d$.

Before we proceed to the discussion of the tensionless
conformal p-brane we  make the following observation. If
we choose the vacuum to be annihilated by the positive
modes, as is the case for the usual string, the
commutators (\ref{d12}) will give
\be
&&\left\langle
0\left |\left[\tilde{L}^{\alpha}_{m_1,\ldots
,m_p},\tilde{L}^{\alpha}_{-m_1,\ldots
,-m_p}\right]\right |0\right\rangle
=2m_{\alpha}\left\langle
0\left
|\tilde{L}^{\alpha}_{0,\ldots,0}\right|0\right\rangle
+m_{\alpha}\tilde{d}^{\alpha}\nonumber\\
&\Rightarrow &
[(d-25-p)m_{\alpha}^3- (d-1-p)m_{\alpha}]\sum_{k_1,
\ldots,k_{\alpha -1},k_{\alpha +1},\ldots,k_p}
\frac{1}{6}=
2m_{\alpha}a^{\alpha}_L+
m_{\alpha}\tilde{d}^{\alpha}\nonumber\\
&\Rightarrow &d=25+p.\label{crst}
\ee
This results agrees with the critical dimension of
$d=27$ for the membrane ($p=2$) which was given in
\cite{umms}. It should be noted here also that
(\ref{crst}) also puts   restrictions on   the
space-time dimension of the tensile bosonic p-brane
theory.

As a side remark we  also note  that the results
obtained here can be  generalized to the case of the
spinning p-brane. Starting of a generalized version of
the action of the tensionless spinning string presented
in \cite{ulbsgt2,ps} we find that the constraints
appropriate for the tensionless p-brane read
\be
\phi^{-1}_{m_1,\ldots ,m_p}&=&\half
\sum_{k_1,\ldots, k_p =-\infty}^{+\infty}p_{m_1 -
k_1,\ldots, m_p - k_p}
\cdot p_{k_1,\ldots ,k_p}=
0,\nonumber\\
L^{\alpha}_{m_1,\ldots
,m_p}&=&\sum_{k_1,\ldots, k_p
=-\infty}^{+\infty} 
\left(-i k_{\alpha}p_{m_1 -
k_1,\ldots, m_p - k_p}\cdot x_{k_1,\ldots
,k_p}\right.\nonumber\\
&&+\half\sum^{N}_{i=1}\left. k_{\alpha}
\psi^{i}_{m_1 -
k_1,\ldots, m_p - k_p}\cdot 
\psi^{i}_{k_1,\ldots ,k_p}\right )=
0,\nonumber\\
S^{i}_{m_1,\ldots ,m_p}&=&\half
\sum_{k_1,\ldots, k_p =-\infty}^{+\infty}p_{m_1 -
k_1,\ldots, m_p - k_p}
\cdot \psi^{i}_{k_1,\ldots ,k_p}=
0.\nonumber
\ee 
where $\psi^{i\mu}$ is the fermionic partner of
$x^{\mu}$, $i=1,\ldots,N$, $N$ being the number of
supersymmetries. 

As it is evident from the previous discussion and the
analysis in \cite{ps}, $\cal{Q}$ will be nilpotent if
the central extensions $\tilde{d}^{\alpha}$ of the
algebra of the extended constraints vanish. Thus we may
 focus only on the commutator
\be
&&\left\langle
0\left |\left[\tilde{L}^{\alpha}_{m_1,\ldots
,m_p},\tilde{L}^{\alpha}_{-m_1,\ldots
,-m_p}\right]\right |0\right\rangle
=2m_{\alpha}\left\langle
0\left
|\tilde{L}^{\alpha}_{0,\ldots,0}\right|0\right\rangle
+m_{\alpha}\tilde{d}^{\alpha}\label{commutator}
\ee
where now
\be
\tilde{L}^{\alpha}_{m_1,\ldots,m_p}&=&
{L}^{\alpha}_{m_1,\ldots,m_p}+\sum_{k_1,\ldots,k_p}\left
[ (k_{\alpha}-m_{\alpha})c_{-k_1,\ldots,-k_p}^{-1}
b_{k_1+ m_1,\ldots,k_p + m_p}^{-1}\right .\nonumber\\
&&+\sum_{\beta
=1}^{p}(
k_{\alpha}c_{-k_1,\ldots,-k_p}^{L_{\beta}}
b_{k_1+ m_1,\ldots,k_p + m_p}^{L_{\beta}}-
m_{\beta}c_{-k_1,\ldots,-k_p}^{L_{\beta}}
b_{k_1+ m_1,\ldots,k_p + m_p}^{L_{\alpha}})\nonumber\\
&&+\sum_{i=1}^{N}(\half
m_{\alpha}-k_{\alpha})\left .
{\gamma}^{i}_{-k_1,\ldots,-k_p}
\beta_{k_1+ m_1,\ldots,k_p + m_p}^{i}\right ],\label{L1}
\ee
and ${\gamma}^{i}_{m_1,\ldots,m_p}$,
${\beta}^{i}_{m_1,\ldots,m_p}$ are 
bosonic ghosts and ghost
momenta corresponding to $S^{i}_{m_1,\ldots,m_p}$.
Plugging (\ref{L1}) into (\ref{commutator})  and
choosing positive modes to annihilate the vacuum we find
the following critical dimension for the spinning p-brane
\be
d=\frac{100+4p-22N}{4+N}.\label{crd}
\ee
This result is the spinning extension of the result
obtained in \cite{umms} and reproduces the results
obtained in \cite{gararual1,ps} for $p=1$. It should
 be noted  also that
(\ref{crd})  puts   restrictions on   the
space-time dimension of the tensile spinning p-brane
theory.

\section {The tensionless conformal p-brane}
\label{conbrane}

In \cite{jiulbs}, the $p=1$ version of (\ref{action}) was
found to be space-time conformally invariant.
 Requiring this symmetry to be a
fundamental symmetry of the theory led  to the
construction of the conformal string, a string with
manifest space-time conformal symmetry \cite{us}.
We find also that the action  (\ref{action}) is
invariant under infinitesimal conformal boosts with
parameter $b^{\mu}$
\be
\delta_{b} X^{\mu}&=&(b\cdot X)X^{\mu}-\half
X^{2}b^{\mu}\cr
\delta_{b} V^{a}&=&-(b\cdot X)V^{a}\nonumber
\ee 
and under infinitesimal conformal dilatations with
parameter $a$
\be
\delta_{a} X^{\mu}&=&aX^{\mu}\cr
\delta_{a} V^{a}&=&-a V^{a}\nonumber
\ee
 The isomorphism $C_{d-1,1}\simeq O(d,2)$ for $d\geq 3$
makes it possible
 to construct a theory in two extra dimensions such that
the previous model corresponds
 to a particular gauge fixing of the latter and the
conformal symmetry is manifest and linearly
 realized
\cite{akul,ISBE,us}. This {\em conformal} tensionless
p-brane action can
 be given by
\be S&=&\int d^{p+1}\sigma
\{V^{a}(\partial_{a}+W_{a}) X^{A}
  V^{b}(\partial_{b}+W_{b})
X_{A} +\Phi
X^{A}X_{A}\},\label{action2}
\ee
where $A=0,\ldots,d+1$ and the new metric has the form
\be
\eta _{AB} = \left( \matrix{\eta_{\mu\nu}\quad  \hfill 0
\quad 0  \cr
 0...0 \quad \hfill 1 \quad 0 \cr 0...0 \quad \hfill 0 -1
\cr}
\right).\nonumber
\ee
$W_{a}$ is the gauge field for scale transformations
and $\Phi $
 is a Lagrange multiplier field.

We can check that by imposing two gauge
 fixing conditions $P^{+}=0$, $X^{+}=1$ the generators of
the Lorentz transformations in the extended
 space become the generators of the conformal group in the
original space. Thus rotations in the
 extended space correspond to conformal 
transformations in
the original space.

Going to the Hamiltonian formulation we  find in exactly
the same manner
 that the Hamiltonian is again a linear combination of 
the
constraints. In addition
 to the original constraints (\ref{ccon1})-(\ref{ccon2})
we will have two new ones
 which in Fourier modes can be written as follows
\be
\phi^{1}_{m_1,\ldots ,m_p}&=&\half
\sum_{k_1,\ldots, k_p =-\infty}^{+\infty}x_{m_1 -
k_1,\ldots, m_p - k_p}
\cdot x_{k_1,\ldots ,k_p}=
0,\label{ccon3}\\
\phi^{0}_{m_1,\ldots
,m_p}&=&\half\sum_{k_1,\ldots, k_p
=-\infty}^{+\infty} 
 p_{m_1 -
k_1,\ldots, m_p - k_p}\cdot x_{k_1,\ldots ,k_p}=
0\label{ccon4}.
\ee  
The constraint algebra with the central extensions
included, for $p>1$, will be given by
\be
\left[
{\phi}^{-1}_{m_1,\ldots ,m_p},\phi^{1}_{n_1,\ldots
,n_p}\right]   &=&
-2i{\phi}^{0}_{m_1
+n_1,\ldots,m_p+n_p}-2ic
\delta_{m_1+n_1}\ldots\delta_{m_p+n_p},\label{excont1}\\
 \left[
{\phi}^{-1}_{m_1,\ldots ,m_p},\phi^{0}_{n_1,\ldots
,n_p}\right]   &=&
-i{\phi}^{-1}_{m_1
+n_1,\ldots,m_p+n_p},\label{excont2}\\
 \left[
{\phi}^{-1}_{m_1,\ldots ,m_p},{L}^{\alpha}_{n_1,\ldots
,n_p}\right]   &=&
(m_{\alpha}-n_{\alpha}){\phi}^{-1}_{m_1
+n_1,\ldots,m_p+n_p},\label{excont3}\\
 \left[
{\phi}^{1}_{m_1,\ldots ,m_p},\phi^{0}_{n_1,\ldots
,n_p}\right]   &=&
i{\phi}^{1}_{m_1
+n_1,\ldots,m_p+n_p},\label{excont4}\\
 \left[
{\phi}^{1}_{m_1,\ldots ,m_p},{L}^{\alpha}_{n_1,\ldots
,n_p}\right]   &=&
(m_{\alpha}+n_{\alpha}){\phi}^{1}_{m_1
+n_1,\ldots,m_p+n_p},\label{excont5}\\
 \left[
{\phi}^{0}_{m_1,\ldots ,m_p},{L}^{\alpha}_{n_1,\ldots
,n_p}\right]   &=&
m_{\alpha}{\phi}^{0}_{m_1
+n_1,\ldots,m_p+n_p}+cm_{\alpha}
\delta_{m_1+n_1}\ldots\delta_{m_p+n_p},\label{excont6}\\
\left[
{L}^{\alpha}_{m_1,\ldots ,m_p},{L}^{\beta}_{n_1,\ldots
,n_p}\right] &=&m_{\beta}L^{\alpha}_{m_1
+n_1,\ldots,m_p+n_p}- n_{\alpha}L^{\beta}_{m_1
+n_1,\ldots,m_p+n_p}\nonumber\\
&&+\half\left
(m_{\beta}d^{\alpha}-n_{\alpha}d^{\beta}\right )
\delta_{m_1+n_1}\ldots\delta_{m_p+n_p}.
\ee  
Comparing the central extensions that appear in this
algebra with the central extensions that appear in the
algebra of the constrains of the conformal string
\cite{us}, we note once again that the former become
more "smooth" then the latter due to the Jacobi
identities. In the case of the conformal string for
example the commutator (\ref{excont1}) reads
\be
\left[  {\phi}^{-1}_{m},{\phi}^{1}_{n}\right]
&=&-2i         
{\phi}^{0}_{m+n}-2(i{d}_{1}+i{d}_{2}m)
\delta_{m+n}.\nonumber
\ee

With the constraint algebra at hand we find the
BRST charge to be
\be
 Q& = &\sum_{k_1,\ldots,k_p}\left [
\phi_{-k_1,\ldots,-k_p}^{-1}c_{k_1,\ldots,k_p}^{-1}+
\right .\sum_{\alpha =1}^{p}
{L^{\alpha}}_{-k_1,\ldots,-k_p}
c_{k_1,\ldots,k_p}^{L_{\alpha}}\nonumber\\
&&\left .+
\phi_{-k_1,\ldots,-k_p}^{0}c_{k_1,\ldots,k_p}^{0}
+
\phi_{-k_1,\ldots,-k_p}^{1}
c_{k_1,\ldots,k_p}^{1}\right ]\nonumber\\
&&+\sum_{k_1,\ldots,k_p}\sum_{l_1,\ldots,l_p}\left [
2i
c_{-k_1,\ldots,-k_p}^{-1}
c_{-l_1,\ldots,-l_p}^{1}
b^{0}_{k_1+ l_1,\ldots,k_p +
l_p}\right .\nonumber\\
&&+
ic_{-k_1,\ldots,-k_p}^{-1}
c_{-l_1,\ldots,-l_p}^{0}
b^{-1}_{k_1+ l_1,\ldots,k_p +
l_p}-i
c_{-k_1,\ldots,-k_p}^{1}
c_{-l_1,\ldots,-l_p}^{0}
b^{1}_{k_1+ l_1,\ldots,k_p +
l_p}\nonumber\\
&&-\half
\sum_{\alpha,\beta
=1}^{p}
k_{\beta}c_{-k_1,\ldots,-k_p}^{L_{\alpha}}
c_{-l_1,\ldots,-l_p}^{L_{\beta}}
b^{L_{\alpha}}_{k_1+ l_1,\ldots,k_p +
l_p}\nonumber\\
&&+\half\sum_{\alpha,\beta
=1}^{p}
l_{\alpha}c^{L_{\alpha}}_{-k_1,\ldots,-k_p}
c_{-l_1,\ldots,-l_p}^{L_{\beta}}
b_{k_1+ l_1,\ldots,k_p +
l_p}^{L_{\beta}}\nonumber\\
&&-
\sum_{\alpha
=1}^{p}
(k_{\alpha}-l_{\alpha})c_{-k_1,\ldots,-k_p}^{1}
c_{-l_1,\ldots,-l_p}^{L_{\alpha}}
b^{1}_{k_1+ l_1,\ldots,k_p +
l_p}\nonumber\\
&&-
\sum_{\alpha
=1}^{p}
(k_{\alpha}-l_{\alpha})c_{-k_1,\ldots,-k_p}^{-1}
c_{-l_1,\ldots,-l_p}^{L_{\alpha}}
b_{k_1+ l_1,\ldots,k_p + l_p}^{-1}\nonumber\\
&&-\sum_{\alpha
=1}^{p}\left .
k_{\alpha}c_{-k_1,\ldots,-k_p}^{0}
c_{-l_1,\ldots,-l_p}^{L_{\alpha}}
b^{0}_{k_1+ l_1,\ldots,k_p +
l_p}\right ].\label{Q1}
\ee  
Again we can check the nilpotency of
 $\cal{Q}$ with the use of
the extended
 constraints. These are given by
\be
\lefteqn{\tilde{\phi}^{-1}_{m_1,\ldots,m_p}=
\phi^{-1}_{m_1,\ldots,m_p}
+\sum_{k_1,\ldots,k_p}\left [-\sum_{\alpha
=1}^{p}
(m_{\alpha}-k_{\alpha})c_{-k_1,\ldots,-k_p}^{L_{\alpha}}
b_{k_1+ m_1,\ldots,k_p + m_p}^{-1}\right .}\nonumber\\
&&\left. +2ic_{-k_1,\ldots,-k_p}^{1}
b_{k_1+ m_1,\ldots,k_p + m_p}^{0}+
ic_{-k_1,\ldots,-k_p}^{0}
b_{k_1+ m_1,\ldots,k_p +
m_p}^{-1}\right ],\label{excon1}\\
\lefteqn{\tilde{L}^{\alpha}_{m_1,\ldots,m_p}=
{L}^{\alpha}_{m_1,\ldots,m_p}+\sum_{k_1,\ldots,k_p}\left
[ (k_{\alpha}-m_{\alpha})c_{-k_1,\ldots,-k_p}^{-1}
b_{k_1+ m_1,\ldots,k_p +
m_p}^{-1}\right.}\label{excon2}\\
 &&+\sum_{\beta
=1}^{p}
\left (k_{\alpha}c_{-k_1,\ldots,-k_p}^{L_{\beta}}
b_{k_1+ m_1,\ldots,k_p + m_p}^{L_{\beta}}-
m_{\beta}c_{-k_1,\ldots,-k_p}^{L_{\beta}}
b_{k_1+ m_1,\ldots,k_p +
m_p}^{L_{\alpha}}\right )\nonumber\\
&&\left. +(k_{\alpha}+m_{\alpha})c_{-k_1,\ldots,-k_p}^{1}
b_{k_1+ m_1,\ldots,k_p + m_p}^{1}+
k_{\alpha}c_{-k_1,\ldots,-k_p}^{0} b_{k_1+
m_1,\ldots,k_p + m_p}^{0}\right ],\nonumber\\
\lefteqn{\tilde{\phi}^{0}_{m_1,\ldots,m_p}=
\phi^{0}_{m_1,\ldots,m_p}
+\sum_{k_1,\ldots,k_p}\left [-\sum_{\alpha
=1}^{p}
m_{\alpha}c_{-k_1,\ldots,-k_p}^{L_{\alpha}}
b_{k_1+ m_1,\ldots,k_p + m_p}^{0}\right.} \nonumber\\
&&\left. +ic_{-k_1,\ldots,-k_p}^{1}
b_{k_1+ m_1,\ldots,k_p + m_p}^{1}-
ic_{-k_1,\ldots,-k_p}^{-1}
b_{k_1+ m_1,\ldots,k_p +
m_p}^{-1}\right ],\label{excon3}\\
\lefteqn{\tilde{\phi}^{1}_{m_1,\ldots,m_p}=
\phi^{1}_{m_1,\ldots,m_p}
+\sum_{k_1,\ldots,k_p}\left [-\sum_{\alpha
=1}^{p}
(m_{\alpha}+k_{\alpha})c_{-k_1,\ldots,-k_p}^{L_{\alpha}}
b_{k_1+ m_1,\ldots,k_p + m_p}^{1}\right.} \nonumber\\
&&\left. -2ic_{-k_1,\ldots,-k_p}^{-1}
b_{k_1+ m_1,\ldots,k_p + m_p}^{0}-
ic_{-k_1,\ldots,-k_p}^{0}
b_{k_1+ m_1,\ldots,k_p + m_p}^{1}\right ].\label{excon4}
\ee
They satisfy the same algebra as the original
constraints. The only thing that remains now is 
to calculate the values
of the
 constants $\tilde{d}^{\alpha}$, $\alpha = 1,\ldots,p$
and
$\tilde{c}$ for the vacuum and ordering introduced 
introduced in the
 previous section. In this case the condition $
p^{A}_{m_1,\ldots,m_p}|0\rangle =0$
together
 with the requirement that the BRST charge (\ref{Q1})
should annihilate the vacuum gives the following
 consistency conditions
$\forall m_1,\ldots,m_p$
\be
  p^{A}_{m_1,\ldots,m_p}|0\rangle 
=c^{1}_{m_1,\ldots,m_p}|0\rangle=
b^{-1}_{m_1,\ldots,m_p}|0\rangle
=0,\nonumber\\ 
  \langle 0|x^{A}_{m_1,\ldots,m_p}  
=\langle 0|c^{-1}_{m_1,\ldots,m_p}= 
 \langle 0|b^{1}_{m_1,\ldots,m_p} =0.\nonumber
\ee
In the same way as was done in the previous section we
find that $\tilde{d}^{\alpha}=0$. We  find the value of
$\tilde{c}$  by calculating  the expectation value of
the commutator
$\left\langle 
0\right|[\tilde{\phi}^{0}_{m_1,\ldots
,m_p},\tilde{\phi}^{L}_{-m_1,\ldots ,-m_p}]\left|
  0\right\rangle $. This gives

\be
\label{fcon2}
 &\left\langle
0\right|[\tilde{\phi}^{0}_{m_1,\ldots
,m_p},\tilde{\phi}^{L}_{-m_1,\ldots ,-m_p}]\left|
  0\right\rangle =m_{\alpha}\left\langle
0\left |\tilde{\phi}^{0}_{0,\ldots,0}\right |
  0\right\rangle
+\tilde{c}m_{\alpha}&\nonumber\\
  &\Rightarrow
0=m_{\alpha}a_{0}+\tilde{c}m_{\alpha}=0&\nonumber\\
  &\Rightarrow 
  \tilde{c}=-a_{0}.&\nonumber
\ee 
But from (\ref{excon3}), for a Hermitian BRST charge
$\cal{Q}$ we will have
\be
\alpha_{0}\equiv \left\langle
0\right|\tilde{\phi}^{0}_{0}\left|
  0\right\rangle =-\frac{i}{4}[d-2]\left(
 \sum_{k_1,\ldots,k_p}1
\right).\nonumber
\ee

The BRST charge is nilpotent when all the constants
$\tilde{d}_{a}$ and $\tilde{c}$ are equal to 0. In
particular we must have
\be
\tilde{c}=0 \Rightarrow d=2.\nonumber
\ee 
Thus we find a critical dimension of $d=2$ for all the
tensionless conformal p-branes. It should be stated here
that the result is valid for $p\neq 0$ since the quantum
theory of the  conformal particle \cite{RMBN} is
consistent in any dimension.

The results obtained here can be easily generalized to
the case of the conformal tensionless spinning p-brane.
Using the techniques presented here and starting from
a generalized version of the action of the conformal
spinning string presented in
\cite{ps}, we   find for the p-brane  a negative
critical dimension 
\be
d=2-2N,\quad \forall p\geq 1,\nonumber
\ee
$N$, being the number of supersymmetries. Again this
result is valid for $p\neq 0$ since for  the conformal
spinning particle the same analysis reveals consistency
 in any dimension, as was first noted in \cite{mart}.

In this paper we have studied the BRST quantization of
the conformal tensionless p-brane and found obstructions
to the quantization except in two space-time dimensions.
Comparing this result with the conformal particle
\cite{RMBN}, which essentially is a model containing the
zero modes of the $\phi$ constraints and has a
quantized theory that is consistent regardless of the
dimensionality of the ambient space, we may conclude
that it is the richness of the state space, a
consequence of the extendedness of the p-brane, which
causes the problems. In two dimensions this extendedness
effectively vanishes. Also, the space-time conformal
group is infinite-dimensional in $d=2$ which is also
expected to give good quantum behavior.

\bigskip
\begin{flushleft} {\bf Acknowledgments:} I would like
to  thank Ulf Lindstr\"om  for many useful discussions.
\bigskip
\end{flushleft}


\begin{thebibliography}{99}

\bibitem{liraspsr} F. Lizzi, B. Rai, G. Sparano and A.
Srivastava, {\it Phys.Lett.} {\bf 182B} (1986) 326.

\bibitem{jiulbs} {J. Isberg,} {U. Lindstr\"om} and B.
Sundborg,
\newblock {\it Phys. Lett.\/}, {\bf 293B} (1992) 321. 


\bibitem{BigT} {J. Isberg,} {U. Lindstr\"om} {B. Sundborg}
and G. Theodoridis,
\newblock {\it Nucl. Phys.} {\bf B411} (1994) 122.

\bibitem{ISBE} J. Isberg, {\it ``Tensionless Strings with
Manifest Space-Time Conformal Invariance''}, Stockholm
University preprint USITP-92-10  (1992).


\bibitem{us}
  H. Gustafsson, U. Lindstr\"om, P. Saltsidis, B. Sundborg
and R.v. Unge,  {\em Nucl. Phys.\/} {\bf B440} (1995)
495.

\bibitem{spectrum}
P. Saltsidis, {\em The mass spectrum of the 2-dimensional
conformal string}, USITP-96-13, hep-th/9609169, to
appear in {\em Phys. Lett.\/} {\bf B}.

\bibitem{HLU}
S. Hassani, U. Lindstr\"{o}m and R. von Unge, {\em Class.
Quantum Grav.\/} {\bf 11} (1994) L79-L85.

\bibitem{ulbsgt1} U. Lindstr\"om, B. Sundborg and G.
Theodoridis,  {\it Phys.Lett.}  {\bf 253B} (1991) 319.


\bibitem{FRAD}
  E.S. Fradkin and G.A. Vilkovisky, {\em Phys. Lett.} {\bf
55B} (1975) 224;
  I.A. Batalin and  G.A. Vilkovisky, {\em Phys. Lett.}
{\bf 69B} (1977) 343.


\bibitem{MarnABRST}
  {R. Marnelius}, {\em Nucl. Phys.} {\bf B294} (1987) 685.

\bibitem{BAZH}
I.A. Bandos and A. A. Zheltukhin {\em Sov. J. Nucl.
Phys.} {\bf 50} (1989) 556.


\bibitem{MarnBRST}
  {S. Hwang} and {R. Marnelius}, {\em Nucl. Phys.} {\bf
B315} (1989) 638.


\bibitem{umms}
U. Marquard and M. Scholl {\em Phys. Lett.\/} {\bf 209B}
(1988) 434.


\bibitem{ulbsgt2} U. Lindstr\"om, B. Sundborg and G.
Theodoridis, {\it Phys.Lett.} {\bf 258B} (1991) 331


\bibitem{ps}
  P. Saltsidis, {\em Nucl. Phys.\/} {\bf B 446} (1995)
286.


\bibitem{gararual1} J. Gamboa, C. Ramirez and M.
Ruiz-Altaba, {\it Phys.Lett.\/} {\bf 225B} (1989) 335.


\bibitem{akul} A. Karlhede and U. Lindstr\"om, {\em
Class.Quant.Grav.\/} {\bf 3} (1986) L73.


\bibitem{RMBN}
  R. Marnelius, {\em Phys. Rev.} {\bf D20} (1979) 2091 ;
  R. Marnelius and B.E.W. Nilsson, {\em Phys. Rev.}
{\bf D22} (1980) 830.


\bibitem{mart}
  {U. M\aa rtensson}, {\em Int. J. Mod. Phys.} {\bf A8}
(1993) 5305.













  





\end{thebibliography}
\end{document}